\documentclass[aps,pra,twocolumn,showpacs,10pt]{revtex4-1}
\usepackage[pdftex]{graphicx}
\usepackage{amsmath,amssymb}
\usepackage{color}

\begin{document}

\title{Experimental quantum decoherence control by dark states of the environment}

\author{Robert St\'{a}rek*$^{1}$}
\affiliation{$^1$Department of Optics, Palack\' y University, 17. listopadu 1192/12,  771~46 Olomouc,  Czech Republic}

\author{Michal Mi\v{c}uda$^1$}
\affiliation{$^1$Department of Optics, Palack\' y University, 17. listopadu 1192/12,  771~46 Olomouc,  Czech Republic}

\author{Ivo Straka$^1$}
\affiliation{$^1$Department of Optics, Palack\' y University, 17. listopadu 1192/12,  771~46 Olomouc,  Czech Republic}

\author{Martina Nov\'{a}kov\'{a}$^1$}
\affiliation{$^1$Department of Optics, Palack\' y University, 17. listopadu 1192/12,  771~46 Olomouc,  Czech Republic}

\author{Miloslav Du\v{s}ek$^1$}
\affiliation{$^1$Department of Optics, Palack\' y University, 17. listopadu 1192/12,  771~46 Olomouc,  Czech Republic}

\author{Miroslav Je\v{z}ek$^1$}
\affiliation{$^1$Department of Optics, Palack\' y University, 17. listopadu 1192/12,  771~46 Olomouc,  Czech Republic}

\author{Jarom\' ir Fiur\' a\v sek$^1$}
\affiliation{$^1$Department of Optics, Palack\' y University, 17. listopadu 1192/12,  771~46 Olomouc,  Czech Republic}

\author{Radim Filip$^1$}
\affiliation{$^1$Department of Optics, Palack\' y University, 17. listopadu 1192/12,  771~46 Olomouc,  Czech Republic}

\begin{abstract}
Coherent interaction of a quantum system with environment usually induces quantum decoherence. However, remarkably, in certain configurations the coherent system-environment 
coupling can be simultaneously  explored to engineer a specific dark state  of the environment that eliminates the decoherence. Here we report on experimental demonstration 
of such protocol for suppression of quantum decoherence by quantum decoherence itself. The protocol is based on indirect control of the environment 
via quantum measurements on quantum probes interacting with the environment prior to the system that should be protected. No direct manipulation with the environment is required to suppress the decoherence.
In our proof-of-principle experiment, we demonstrate protection of a single qubit coupled to another single qubit. We implement the required quantum circuits with linear optics 
and single photons, which allows us to maintain very high degree of control and flexibility in the experiment.
Our results clearly confirm the decoherence suppression achieved by the protocol and pave the way to its application 
to other physical platforms. 
\end{abstract}

\maketitle

\section*{Introduction}
Noise and decoherence represent significant obstacles in designing large-scale quantum information processing devices or 
achieving long-distance quantum communication. Several ingenious methods have been developed to overcome these detrimental 
effects and enable scalable quantum information processing. The most important techniques to eliminate noise and decoherence include 
quantum error correction \cite{Shor1995,Steane1996,Calderbank1996,Nielsen2000,Yao2012,Bell2014}, 
entanglement distillation \cite{Bennett1996,Deutsch1996,Pan2003,Hage2008,Dong2008,Takahashi2010,Kurochkin2014,Abdelkhalek2016}, 
quantum repeaters \cite{Briegel1998,Duan2001,Chou2007,Yuan2008} or encoding information into decoherence-free 
subspaces \cite{Kwiat2000,Kohout2008,Lidar2014}. The noise and decoherence are commonly modeled by the coherent interaction of the system with environment \cite{Zurek2003} 
that is usually assumed to be large and consisting of an infinite number of elementary constituents (such as quantum harmonic oscillators or qubits). 
However, as the technology advances, the experimentally implemented quantum processors develop towards a regime where the decoherence may be predominantly caused by 
coupling to a finite and possibly small number of nearby quantum objects forming an effective local environment, for example in quantum dots \cite{Hartke2018}, 
superconducting circuits \cite{Lisenfeld2015}, many body systems \cite{Rubio-Abadal2018}, or nitrogen-vacancy centers 
\cite{Kalb2016, Abobeih2018, Hernandez-Gomez2018, Gangloff2019}. Such small local environments induce specific decoherence effects \cite{Zurek2003, Zhao2011, Zwolak2016, Luitz2017}.
It is thus important to investigate efficient and practically feasible techniques for decoherence suppression in such setting 
\cite{Taylor2003, Yao2007, Cappellaro2009, Bluhm2010, Kolkowitz2012, Taminiau2012, Liu2013, Hansom2014, Cramer2016}.

If the coherent interaction with the effective environment has a suitable form, then one can attempt to switch off this interaction by preparing the environment 
in a suitable dark state, which decouples it from the system qubits that we want to protect. Although the environment is assumed to be directly inaccessible, 
it can be influenced {\em{indirectly}} via measurements on the system qubits after their interaction with the environment. One thus utilizes the remote state preparation 
protocol \cite{Bennett2001} that benefits from quantum correlations between the system and the environment that were established uniquely by their interaction responsible for the decoherence. 
Such protocol has been recently theoretically proposed in Ref. \cite{Roszak2015}, where it was illustrated on the proof-of-principle example of a single system qubit coupled to an environment 
represented by a single quantum harmonic oscillator. 

In the present paper, we report on the proof-of-principle experimental demonstration of this idea for protection against decoherence.  
In our test, to keep the protocol simple, both the system and the environment are represented by a single qubit. We encode qubits into path and polarization degrees of freedom of single photons 
and implement the required three-qubit quantum logic circuit with linear optics. We comprehensively characterize the decoherence suppression protocol 
by quantum state and quantum process tomography. Our results clearly verify the practical utility of decoherence suppression via \emph{indirect} environment control, 
thus opening way for its application to solid-state platforms of quantum technology.

\begin{figure*}[t!]
\centerline{\includegraphics[scale=0.95]{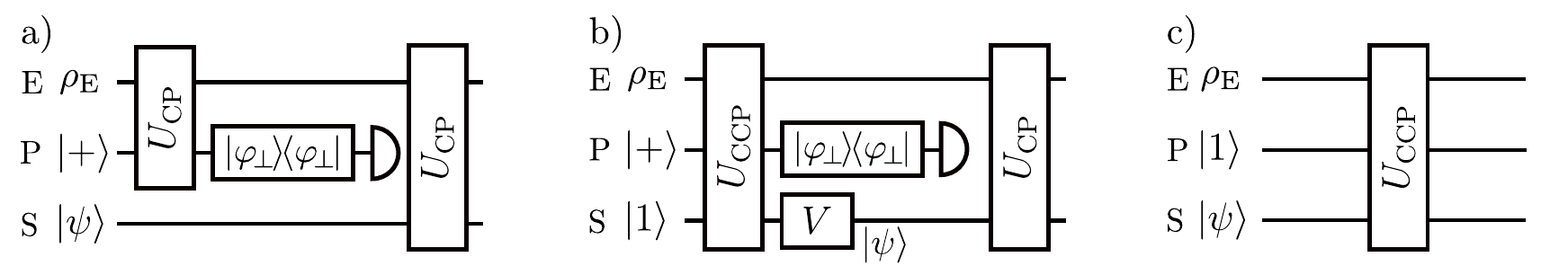}}
\caption{
{\bf a)} Quantum circuit implementing the proof-of-principle test of decoherence suppression protocol. The circuit involves three qubits: environment (E), probe (P) and signal (S). 
Qubits E and P are coupled by $U_{\mathrm{CP}}$ gate. Projection of the qubit P onto the state $|\varphi_{\perp}\rangle$ heralds successful preparation of the qubit E in the dark state $|0\rangle$. 
The qubit S is then not altered by interaction with qubit E. 
{\bf b)} Experimentally implemented quantum circuit, equivalent to circuit (a). Here qubit S is initially prepared in a fixed state $|1\rangle$ and all three qubits E, P and S are coupled by $U_{\mathrm{CCP}}$ operation. 
An arbitrary pure state $|\psi\rangle$ of the qubit S is then prepared with the use of a single-qubit gate $V$. {\bf c)} Reference quantum circuit for observing decoherence effects of the environment. 
This circuit is equivalent to the two-qubit gate $U_{\mathrm{CP}}$ acting on qubits E and S. 
}
\label{fig:1}
\end{figure*}

\section*{Results and discussion}

\subsection*{Theory} 
Let us begin with a theoretical description of the protocol. We consider a simple yet practically important and ubiquitous coupling of initially uncorrelated system (S) and environmental (E) qubits 
via unitary controlled phase shift operation,
 \begin{equation}
 U_{\mathrm{CP}}(\varphi)=\exp(i\varphi |1\rangle \langle 1|_{\mathrm{S}} \otimes |1\rangle \langle 1|_{\mathrm{E}}),
 \label{eq:UCP}
 \end{equation}
 where $|0\rangle$ and $|1\rangle$ denote two basis states of a qubit and $\varphi\in[0,2\pi)$ specifies the coupling strength. 
This system-environment coupling  reduces the coherence between the two computational basis states of the system qubit. 
 Specifically, $ \rho_{\mathrm{S},01} \rightarrow q \rho_{\mathrm{S},01}$, where $\rho_{\mathrm{S},01}=\langle 0 |\rho_\mathrm{S} |1\rangle$, 
 $q=p_{0,\mathrm{E}}+e^{i\varphi}p_{1,\mathrm{E}}$, and $p_{j,\mathrm{E}}=\langle j|\rho_\mathrm{E}|j\rangle$
is the probability that the environmental qubit is in state $|j\rangle_{\mathrm{E}}$. Here $\rho_\mathrm{S}$ and $\rho_\mathrm{E}$ denote initial density matrices of system and environmental qubit.   
It holds that $|q|<1$ unless $\varphi=0$, $p_{0,\mathrm{E}}=1$ or $p_{1,\mathrm{E}}=1$. 
  
 The interaction  between the system and the environment can be switched off provided that 
the environmental qubit, which is not directly accessible, can be prepared in state $|0\rangle_{\mathrm{E}}$. This can be achieved by the following protocol.  
Before the system qubit is used for information encoding, it is prepared in a pure superposition state
 $|+\rangle_{\mathrm{S}}=\frac{1}{\sqrt{2}}(|0\rangle+|1\rangle)_{\mathrm{S}}$, and after interaction with the environment it is projected onto state
\begin{equation}
|\varphi_\perp\rangle_{\mathrm{S}}= \frac{1}{\sqrt{2}}\left(|0\rangle-e^{i\varphi}|1\rangle\right)_{\mathrm{S}}.
\end{equation}
It is easy to verify that $_{\mathrm{S}}\langle \varphi_\perp|U_{\mathrm{CP}}(\varphi)|+\rangle_{\mathrm{S}}|1\rangle_{\mathrm{E}}=0$ hence successful projection of the system qubit 
 onto state $|\varphi_\perp\rangle_{\mathrm{S}}$ heralds preparation of the environment in a dark state $|0\rangle_{\mathrm{E}}$.

The success probability of the protocol reads 
 \begin{equation}
 P_{S}=p_{0,\mathrm{E}}|\langle\varphi_\perp|+\rangle|^2=p_{0,\mathrm{E}}\sin^2\frac{\varphi}{2}.
 \label{PSdef}
 \end{equation}
 Therefore, unless the environmental qubit is initially in a pure state $|1\rangle_{\mathrm{E}}$, 
 we can always conditionally switch the decoherence off before the system qubit is used for some application.
Notably, the success probability $P_{S} $ increases with increasing interaction strength up to maximum at $\varphi=\pi$ and vanishes in the limit $\varphi \rightarrow 0$. 
In this limit, the environment affects the system qubit only very weakly and thus the output states of the system qubit corresponding to the environmental states 
$|0\rangle_{\mathrm{E}}$ and $|1\rangle_{\mathrm{E}}$ become almost indistinguishable. By contrast, for $\varphi=\pi$ those two states become orthogonal.

If the system qubit is not projected on the desired state $|\varphi_\perp\rangle_{\mathrm{S}}$ but on its orthogonal counterpart $|\varphi\rangle_{\mathrm{S}}=\frac{1}{\sqrt{2}}(|0\rangle+e^{i\varphi} |1\rangle)_{\mathrm{S}}$,
we may attempt to repeat the remote state preparation procedure until it succeeds. This assumes that the system qubit can be repeatedly refreshed and prepared in state $|+\rangle_{\mathrm{S}}$. 
The success probability after up to $N$ repetitions of the protocol is given by $P_{S,N}=p_{0,\mathrm{E}}\left[1-\cos^{2N}(\varphi/2)\right]$.
This probability can be further increased if the environment could thermalize to its initial state after each unsuccessful attempt. 
In such case, the overall probability of success after up to $N$ repetitions reads $P_{S,N}=1-(1-P_S)^N$ where $P_S$ is the single-shot success probability given by Eq. (\ref{PSdef}). 
Asymptotically, we can thus reach deterministic preparation of the dark state $|0\rangle_{\mathrm{E}}$, unless the environment is initially in state $|1\rangle_{\mathrm{E}}$.
We note that in contrast to decoherence suppression via weak measurements and quantum measurement reversal  \cite{Korotkov2010,Lee2011,Kim2012} which relies on an application of suitable quantum filters to the measured qubit, our goal here is to perform an \emph{indirect} projective measurement on the environmental qubit, not a weak measurement. Also, the target of our protocol is the environment and not the decohered system qubit that just serves as a proxy to indirectly control the environment.

So far we have assumed that the coupling strength $\varphi$ is known. This allows us to decouple the environment from the system in a single shot. 
However, in practice, $\varphi$ may be unknown or may even fluctuate in time. 
 In such case one can conservatively choose to project the output state of the signal qubit onto state $|+\rangle_{\mathrm{S}}$. 
 Although this generally does not exactly prepare the environmental qubit in state $|0\rangle_{\mathrm{E}}$, 
 it reduces the probability $p_{1,E}$ with respect to $p_{0,\mathrm{E}}$. If we define the population ratio 
 $R_{\mathrm{E}}=p_{1,\mathrm{E}}/p_{0,\mathrm{E}}$, then after a single implementation of the protocol with projection onto $|+\rangle$ we obtain 
 \begin{equation}
 R_{\mathrm{E}}'=R_{\mathrm{E}} \cos^2\frac{\varphi}{2}.
 \end{equation}
 Hence we obtain reduction by a factor of $\cos^2(\varphi/2)$ which is smaller than 1 for all $0<\varphi < 2\pi$. If the protocol can be repeated, 
 then after $N$ successful iterations we obtain exponentially strong reduction of the undesired state $|1\rangle_{\mathrm{E}}$ by factor $\left(\cos^2\frac{\varphi}{2}\right)^{N}$.
The qubit S can be therefore asymptotically perfectly protected even without prior knowledge of the coupling strength $\varphi$. 
Note that this protocol is not limited to the interaction~(\ref{eq:UCP}). Analogical procedure can be applied for any unitary coupling
$\exp(i\varphi \sigma_S \otimes A_E)$, where $\sigma_S$ is any Pauli matrix and $A_E$ is an Hermitian operator of an arbitrarily complex environment 
\cite{Taylor2003, Yao2007, Cappellaro2009, Bluhm2010, Liu2013, Hansom2014}.

 \begin{figure*}[!t!]
 	\centerline{\includegraphics[width=0.7\linewidth]{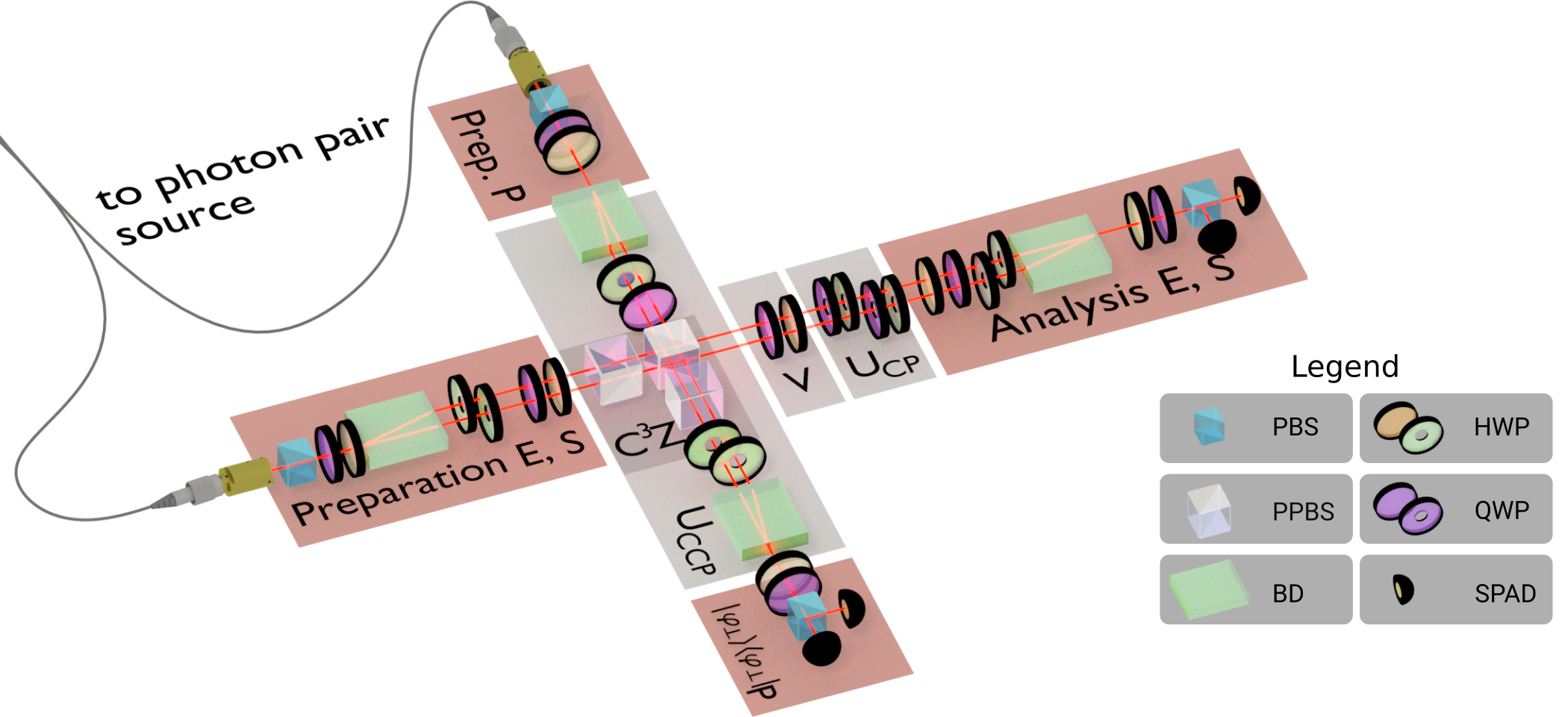}}
 	\caption{ Linear optical setup implementing the quantum circuit of Fig.~1b. This multipath optical interferometer is built from calcite beam displacers (BD), 
 		wave plates (HWP and QWP), and partially polarizing beam splitters (PPBS). Qubits are encoded into path and polarization of single photon. Each output qubit can be measured 
 		in an arbitrary basis with the use of wave plates, polarizing beam splitters (PBS) and avalanche photodiodes (SPAD). Note that the calcite beam displacers 
 		serve as convertors between path and polarization encoding.}
 	\label{fig:2}
 \end{figure*}

\begin{figure*}[t]
\includegraphics[scale=1]{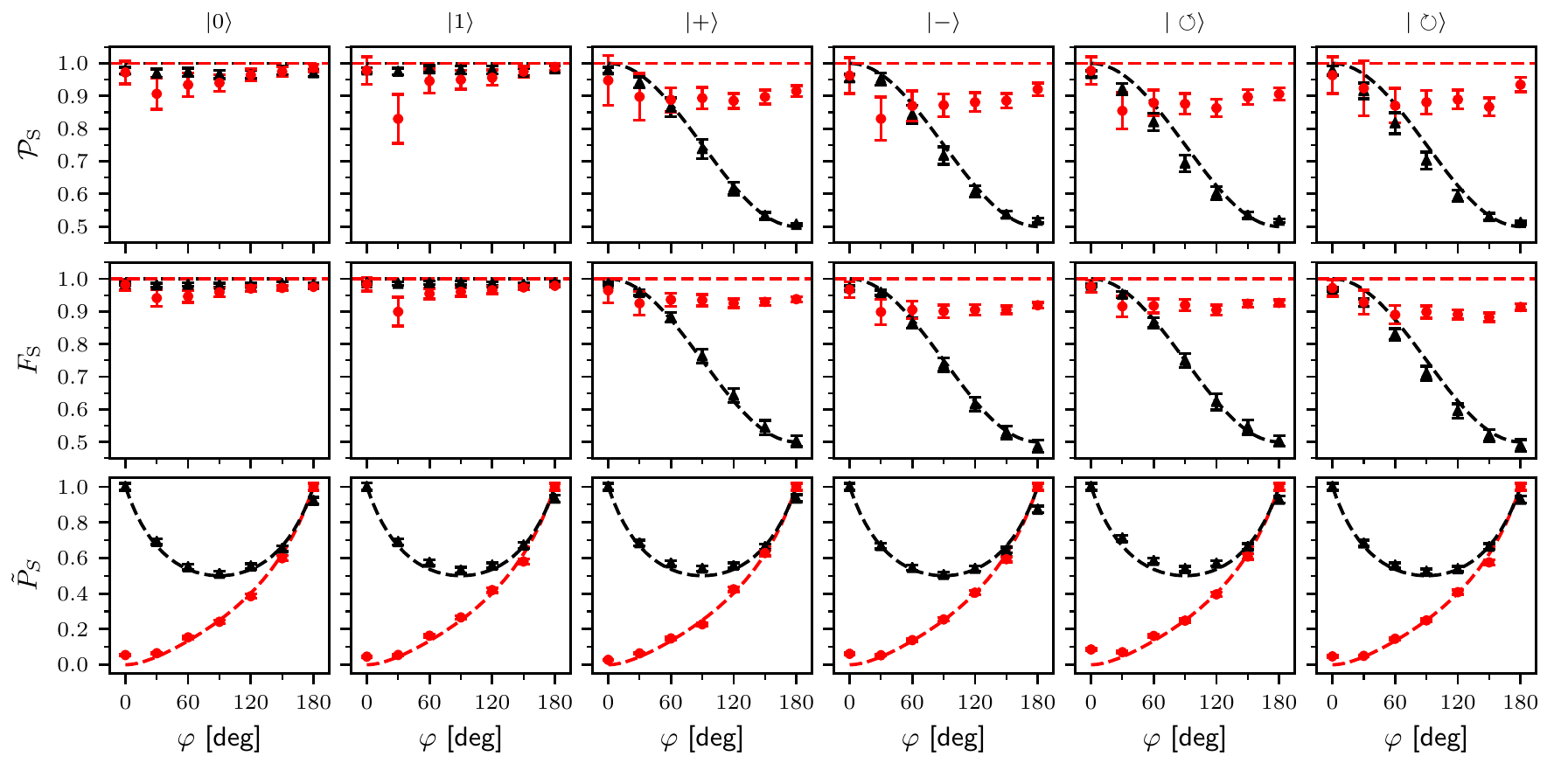}
\caption{Experimental verification of decoherence suppression. The figure shows the output system qubit purity $\mathcal{P}_{\mathrm{S}}$ 
(upper row) and fidelity $F_{\mathrm{S}}$ (middle row), as well as the normalized success probability of the protocol $\tilde{P}_{S}$
(bottom row) in dependence on the system-environment coupling strength $\varphi$. 
The results are plotted for six different input states of signal qubit that form three mutually unbiased bases. Red dots represent 
experimental data, red dashed lines indicate predictions of theoretical model. For comparison, black triangles and lines show results 
for a reference scheme where the system qubit is unprotected and interacts with the environment. The environment qubit is initialized
in a maximally mixed state. The error bars represent three standard deviations. Note that the plotted success probability $\tilde{P}_S$ 
includes the success probability of implementation of the linear optical CCP gate.}
\label{fig:3}
\end{figure*}

\subsection*{Experiment} 
In the experiment, we aim to verify that the interaction with the environment was switched off by the above-described procedure. 
We, therefore, utilize two different system qubits, probe P and signal S. 
The probe serves for the prior control of the quantum state of the environment. Signal qubit is then used to test the success of the decoherence suppression protocol. 
The protocol thus requires two controlled-phase gates $U_\mathrm{CP}$, one between the probe and environmental qubits and the other acting on the signal and environmental qubits, 
as depicted in Fig.~\ref{fig:1}a.

To achieve  high coherence and controllability, we encode qubits into path and polarization of single photons and implement the quantum gates with linear optics \cite{Knill2001,Kok2007}. 
In this approach, it is actually more straightforward to
 implement a sequence of a three-qubit controlled-controlled-phase gate 
\begin{equation}
 U_{\mathrm{CCP}}= \exp\left(i \varphi |1\rangle\langle1|_{\mathrm{P}}  \otimes |1\rangle\langle1|_{\mathrm{S}}  \otimes |1\rangle\langle 1|_{\mathrm{E}} \right), 
 \end{equation}
 and a two-qubit controlled-phase gate $U_{\mathrm{CP}}$.
With signal qubit initially prepared in state $|1\rangle_{\mathrm{S}}$, the $U_{\mathrm{CCP}}$ gate effectively acts as $U_{\mathrm{CP}}$ gate for the remaining qubits E and P.
The experimentally implemented circuit depicted in Fig.~\ref{fig:1}b is therefore fully equivalent to the circuit in Fig.~\ref{fig:1}a.
We also utilize the quantum circuit depicted in Fig.~\ref{fig:1}c that serves as a reference to directly observe the effect of unsuppressed decoherence.

Our experimental setup is depicted in Fig.~\ref{fig:2}. 
Qubits are encoded into polarization and path degrees of freedom of two single photons generated in the process of spontaneous parametric down-conversion. 
Specifically, the environmental qubit is represented by path of the signal photon, the signal qubit is encoded into polarization of the signal photon and the probe qubit is encoded 
into polarization of the idler photon. Inherently stable Mach-Zehnder interferometers formed by calcite beam displacers are utilized to support path-encoded qubits. 
Polarizing beam splitters, wave plates, partially polarizing beam splitters, and avalanche photodiodes serve for qubit state preparation, manipulation, and detection.  
With our encoding, the two-qubit controlled-phase gate $U_{\mathrm{CP}}$ can be implemented deterministically by a sequence of wave plates \cite{Fiorentino2004}. 
The three-qubit gate $U_{\mathrm{CCP}}$ that couples the two photons is realized by a combination of two-photon and single-photon interference 
and postselection on observation of two-photon coincidence counts between detection blocks P and S \cite{Okamoto2005,Langford2005,Kiesel2005,Lanyon2009,Lemr2011,Micuda2013,Micuda2015}. 
The implementation of the three-qubit CCP gate is thus probabilistic and the probability adds to the losses in the experimental setup. 
When the experimental setup is configured to maximal transmittance ($\varphi = 0$), we detect approximately 300 two-photon coincidences per second at the output of the photonic circuit.
The Methods section contains a detailed analysis of the gate operation and performance as well as more details on the experimental setup.

We note that in our proof-of-principle experiment we have direct access to the environmental qubit, but this access is only utilized to prepare a physically well motivated initial state of the environment.
Specifically, we consider the worst-case scenario and the environmental qubit is initially prepared in a maximally mixed state.  
Similarly, measurement of the output environmental qubit is performed only due to the way we technically implement our linear optical quantum gate that operates in the coincidence basis \cite{Kok2007}. 
However, we effectively trace over the environmental qubit by summing the measured coincidence counts over all outcomes of measurements on this qubit. To maximize the homogeneity of this procedure, 
we  sum over outcomes of sequential projective measurements in three mutually unbiased bases.   
The environment is thus effectively discarded and the information from measurement on the environmental qubit is erased and not used in the implemented protocol.

\subsection*{Experimental results}
We have tested the performance of the protocol for various interaction strengths $\varphi$ and for six signal qubit states from the set 
$\mathcal{S} = \{ |0\rangle,\, |1\rangle,\, |+\rangle,\, |-\rangle,\, |\circlearrowleft\rangle=\frac{1}{\sqrt{2}}{(|0+i|1\rangle)},\, |\circlearrowright\rangle=\frac{1}{\sqrt{2}}(|0-i|1\rangle) \}$ 
that contains three mutually unbiased bases. 
The environmental qubit is prepared in a maximally mixed state $\rho_{\mathrm{E}}=\frac{1}{2}I$, which is generated as an equal mixture of the six states from $\mathcal{S}$, 
and the probe qubit is initially prepared in pure input state $|+\rangle_{\mathrm{P}}$. The probe qubit was measured in the superposition basis $\frac{1}{\sqrt{2}}(|0\rangle \pm e^{i\varphi}|1\rangle)$ and projection onto 
$|\varphi_\perp\rangle=\frac{1}{\sqrt{2}}(|0\rangle -e^{i\varphi}|1\rangle)$ heralded decoupling of the environmental qubit. We have performed full quantum state tomography of the output signal 
qubits which allowed us to completely characterize the performance of our protocol. 

The results are depicted in Fig.~\ref{fig:3}. The red dots represent experimental data and the red dashed lines indicate predictions of an ideal theoretical model of our quantum circuit, which does not consider any imperfections.
For each of the six signal qubit states we plot the dependence  of the output state purity $\mathcal{P}_{\mathrm{S}}=\mathrm{Tr}[\rho_{\mathrm{S}}^2]$ 
and fidelity $F_{\mathrm{S}}=\langle \psi|\rho_{\mathrm{S}}|\psi\rangle$ on the phase shift $\varphi$. 
We also provide plot of the relative success probability of our protocol $\tilde{P}_{S}(\varphi)=P_S(\varphi)/P_S(\pi)$. 
Since the total success probability of the protocol $P_S(\varphi)$ is influenced by various loss factors in the optical setup, it is hard to estimate. We therefore utilize the relative success probability 
that can be reliably determined from the collected data as a ratio of coincidence counts.  This approach assumes that our source produces constant number of photon pairs per second on average, 
which was confirmed by independent measurement. Note that besides the bona-fide success probability of the protocol, $\tilde{P}_{S}$ includes also the success probability of the conditional
three-qubit CCP gate which depends on $\varphi$, $P_{\mathrm{CCP}}=\frac{1}{9 + 9|\sin\varphi|}$. 
The error bars were determined using Monte Carlo (bootstrapping) method, described in Methods section. 
For completeness, we have also performed measurements for pure initial state of the environmental qubit, $|+\rangle_{\mathrm{E}}$.  
The results are provided in the Supplementary Material and plotted in Fig.~S1.

\begin{figure}[t]
\centering
\includegraphics[scale=1]{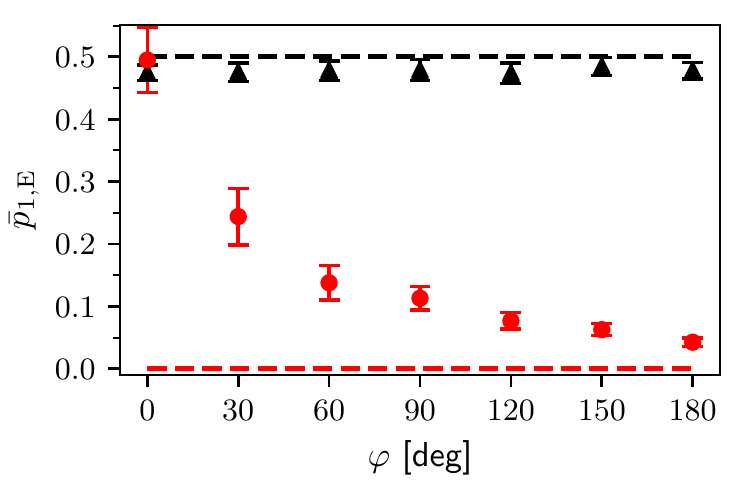}
\caption{Mean population $\bar{p}_{1,\mathrm{E}}$ after decoherence suppression (red dots) and without decoherence suppression (black triangles). 
In both cases, the mean value is calculated over all six tested states of signal qubit. The error-bars show three standard deviations.}
\label{fig:4}
\end{figure}

The measured fidelities in Fig.~\ref{fig:3} are smaller than $1$, which indicates presence of residual decoherence. Since in our experiment we have access to the output environmental qubit, we have performed its  tomographic characterization 
to check whether we have indeed prepared the environment in the dark state.  From the reconstructed density matrix we have determined the residual population $p_{1,\mathrm{E}}$ 
of state $|1\rangle_E$ at the output of the protocol. The populations $p_{1,\mathrm{E}}$ are averaged over all 6 investigated input signal states and plotted as $\bar{p}_{1,\mathrm{E}}$ in Fig.~\ref{fig:4} as red dots. 
Experimentally obtained $\bar{p}_{1,\mathrm{E}}$ are positive, although they should be ideally zero for every coupling strength. 
The residual population of $|1\rangle_E$ causes partial coupling of the signal and environment qubits which reduces purity and fidelity of the output signal qubit.

Figure~\ref{fig:4} indicates that $p_{1,E}$ increases with decreasing coupling strength $\varphi$. 
This can be easily understood, since the weaker is the coupling strength, the lower is the success probability $P_{S}$, because we are filtering out most of the photons. In the limit of weak coupling, 
the remote control of environmental qubit thus becomes very sensitive to experimental imperfections, such as partial distinguishability of the single photons, 
finite extinction ratio of polarizing components, wave-plate retardation and calibration errors, or interferometric phase instability. 
The protocol is thus most efficient in the strong decoherence regime, when the decoherence suppression is most desirable, while it is not very effective for weak system-environment coupling,
when the decoherence effect is also weak. 
The resulting discrepancy between the ideal theoretical prediction and the observed experimental values in Fig.~\ref{fig:3} is the consequence of interplay 
between the coupling strength $\varphi$ and the residual population $p_{1,\mathrm{E}}$ caused by experimental imperfections.


As a reference and benchmark, we have also determined the change of the signal qubit induced by coupling to the environment when the protocol is not implemented and the probe qubit is not used. 
A quantum circuit of this reference measurement is shown in Fig.~\ref{fig:1}c. The probe qubit was set to state $|1\rangle$ and the signal qubit was prepared in the state $|\psi\rangle$ already 
at the input of the three-qubit CCP gate. The other gates were set to identity operations, and the output state of the signal qubit was again characterized by quantum state tomography. 
The experimental results for unsuppressed decoherence are plotted in Fig.~\ref{fig:4} as black triangles together with predictions from an ideal theoretical model represented by black dashed lines. 
It can be clearly seen that coupling to the environment causes dephasing of the signal qubit and reduces purity and fidelity of all superposition states. The relative success probability 
for this reference measurement is defined as $\tilde{P}_S(\varphi)=P_{S}(\varphi)/P_{S}(0)$ since the largest $P_{S}$ is observed for $\varphi=0$. 
While the quantum logic circuit in Fig. 1c as such does not involve any conditioning, our implementation of the CCP gate does.
The observed explicit dependence of $\tilde{P}_S$ on $\varphi$ thus essentially represents the dependence of the relative success probability of the CCP gate on $\varphi$.

\begin{figure}[t]
\centerline{\includegraphics[scale=1]{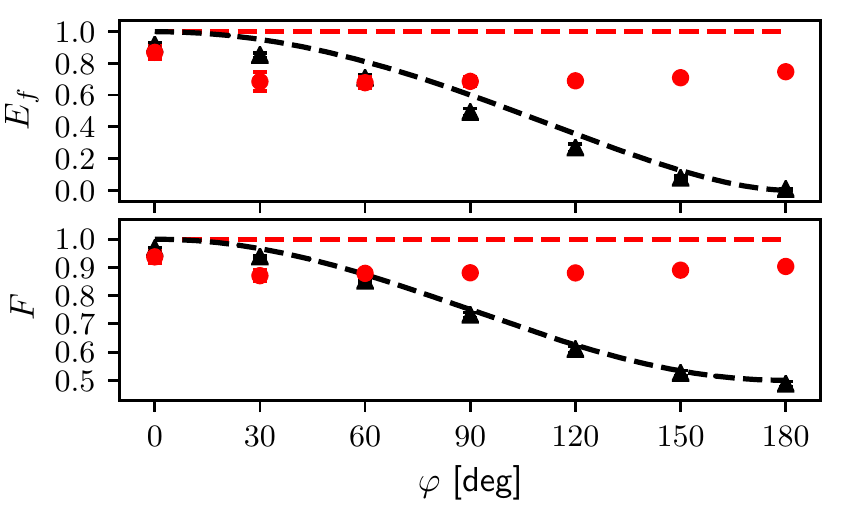}}
\caption{Parameters characterizing the single-qubit quantum channel representing decoherence of signal qubit. Entanglement of formation $E_f$ of the output signal qubit with an external qubit for an input maximally 
entangled Bell state and fidelity $F$ of the channel with the ideal identity channel are plotted for the decoherence suppression protocol (red dots) and for reference measurement with decoherence (black triangles). 
Respective dashed lines show prediction from the ideal theoretical model. The error bars represent three standard deviations.}
\label{fig:5}
\end{figure}

The decoherence of the signal qubit can be comprehensively described by the formalism of quantum channels. 
According to the channel-state duality \cite{Jamiolkowski1972, Choi1975}, a single-qubit channel $\mathcal{D}$ can be represented by applying $\mathcal{D}$ to one part of a maximally entangled two-qubit Bell state 
 $|\Phi^{+}\rangle=(|00\rangle+|11\rangle)/\sqrt{2}$, which yields a quantum process matrix $\chi$ isomorphic to $\mathcal{D}$. 
From the collected tomographic data, we have reconstructed the process matrix $\chi$ \cite{Jezek2003} 
and evaluated its entanglement of formation \cite{Wooters1998} $E_f$ and fidelity $F$ with the Bell state $|\Phi^{+}\rangle$. 
The fidelity indicates how close the channel is to the ideal identity channel, and the entanglement of formation quantifies how entanglement 
of the signal qubit with an external qubit is affected by the decoherence.  Figure~\ref{fig:5} shows the comparison between the cases with and without decoherence suppression protocol. 
Importantly, the decoherence suppression technique helps to preserve entanglement in cases where the entanglement would be otherwise significantly reduced or even completely lost.
Our experiment thus clearly and unambiguously demonstrates the usefulness of the decoherence suppression protocol for protection of quantum information. Advantageously, 
this protocol uses {\em only} the natural interaction between the system and environment to switch off the destructive impact of the environment.
In other words, the very same system-environment interaction that causes the decoherence can turn the environment into dark state, and no additional control of the environment is needed.

\section*{Conclusions and outlook}
In summary, we have demonstrated quantum decoherence suppression  protocol that is based on \emph{indirect} engineering of the quantum state of environment through the decoherence process. 
A crucial feature of the demonstrated protocol is that it involves only a measurement on the system qubit subject to decoherence and  no direct manipulation with the environment or its measurement is required. 
The method is applicable whenever a dark state of the environment can be identified and the quantum correlations between signal and environment established by their interaction 
have suitable structure to allow indirect manipulation of the environmental state. The reported protocol thus provides a useful addition to the toolbox of techniques for fighting noise 
and decoherence in emerging quantum processors where noise and errors may stem 
from coupling to a small well-defined quantum system. Beyond the already proposed application of this protocol to quantum dots \cite{Roszak2015}, 
the method can also be useful for decoherence suppression in many other physical platforms such as trapped ions \cite{Lv2018, Fluehmann2019}, superconducting circuits \cite{Yoshikara2019, Langford2017, Braumueller2017}, 
or nanomechanical oscillators \cite{LaHaye2009, Pirkkalainen2013, Gustafsson2014}.

\begin{figure*}[!t!]
\centerline{\includegraphics[width=0.99\linewidth]{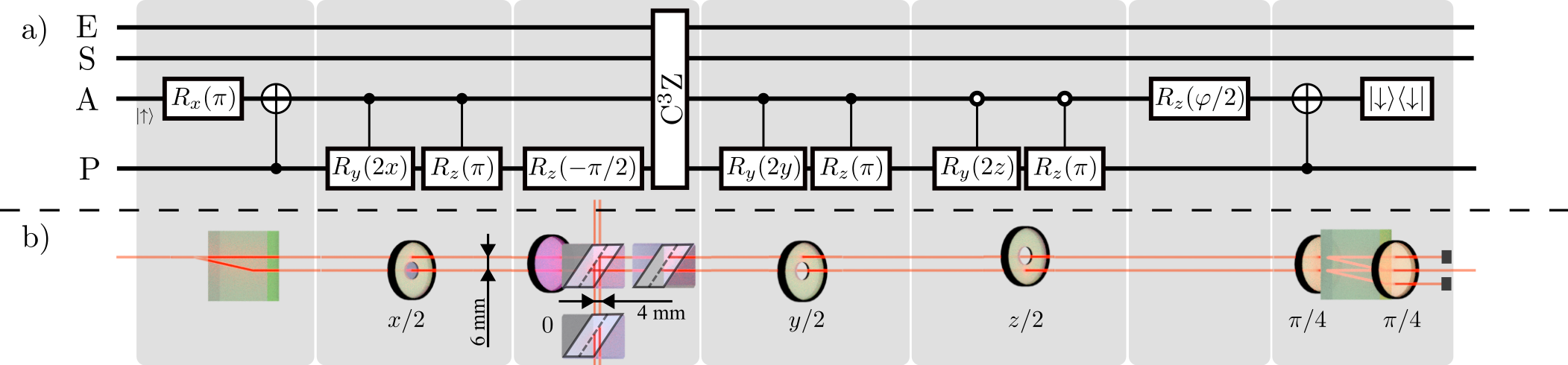}}
\caption{Circuit diagram of the controlled-controlled-phase gate and its respective linear-optical implementation. Empty circles in controlled-operation symbols denote that the operations 
are applied on target qubit when the control qubit is in state $|0\rangle$ ($|\downarrow\rangle$). Half-wave plates in right-most block are depicted for simplicity. In the actual experiment, 
they are omitted and its function is taken into account by appropriate setting of angles $y$, $z$, and angles of wave plates in the analysis of the probe qubit. }
\label{fig:6}
\end{figure*}

\section*{Methods}

\subsection*{Details of experimental setup}
In our experiment, we utilize pair of time-correlated photons generated in the process of spontaneous parametric down-conversion in a nonlinear crystal pumped by a diode laser. 
The resulting signal and idler photons at $810$ nm have orthogonal polarization and are spatially separated by a polarizing beam splitter, coupled into single-mode fibers, 
guided into input ports of our main setup shown in Fig.~2, and there released into free space. The coherence length of the generated photon pairs is approximately 100~$\mu$m. 
We can encode two qubits into each photon, one in polarization and the other in path in an optical interferometer. The computational basis states $|0\rangle$ and $|1\rangle$ 
correspond to horizontal and vertical linear polarization states $|H\rangle$ and $|V\rangle$ or to the propagation  of a photon in the top or bottom arm of an interferometer ($|\uparrow\rangle$, $|\downarrow\rangle$). 
Three qubits are directly utilized in our experiment to test the decoherence suppression protocol, and the spatial degree of freedom of the idler photon allows us to arbitrarily set the coupling strength $\varphi$ of the three-qubit CCP gate, as described below.

The two Mach-Zehnder interferometers (MZI) formed by calcite beam displacers are designed such that they exhibit different spacing between the two interferometer arms.
Specifically, the distance between the two spatial paths is 4~mm for the signal MZI and 6 mm for the idler MZI. This ensures that photons traveling in spatial modes $|0\rangle$ do not 
overlap on the central partially polarizing beam splitter (PPBS), see Fig.~6. The closeness of the two spatial modes provides the inherent phase stability of the interferometers \cite{Starek2018}, 
but brings the experimental inconvenience of placing polarizing elements into a single interferometer arm.
Moreover, due to the short coherence length, each component in a single MZI arm has to be compensated by placing another element into the other arm. We solve those inconveniences 
with ring-shaped wave plates, which can easily address a single arm of the MZI using standard optical mounts. When needed, self-compensation of such wave plates is achieved 
by embedding a glass plate in the center of the ring wave plate.

\subsection*{Controlled-controlled-phase gate}

Here we describe the experimental implementation of $\mathrm{U}_{\mathrm{CCP}}$ which is a crucial part of the experiment. The quantum circuit diagram in Fig.~\ref{fig:6} 
depicts how we constructed the three-qubit $\mathrm{U}_{\mathrm{CCP}}$ gate with an arbitrarily tunable phase shift $\varphi$ from a fixed four-qubit gate and several fixed and tunable two-qubit and single-qubit gates,  
and how we implemented it using linear optics. 
The auxiliary qubit A utilized in our circuit \cite{Lanyon2009} is the path qubit of the idler photon. 
The two-qubit CNOT gates are effectively implemented 
by the calcite beam displacers that couple polarization and path degrees of freedom. Single-qubit rotations $\mathrm{R}_k(\xi)=\exp(i\frac{\xi}{2}\sigma_k)$ as well 
as two-qubit controlled rotations $\mathrm{CR}_k$ are mostly implemented with the help of 
wave plates addressing either both or only a single arm of the Mach-Zehnder interferometer. Here $\sigma_k$, $k = x,\,y,\,z$, denote Pauli matrices.   
An exception is the single-qubit phase shift operation $\mathrm{R}_z(\varphi/2)$ that is achieved by tilting of the second calcite beam displacer.
A particular conditional phase shift $\varphi$ of the $\mathrm{U}_{\mathrm{CCP}}$ gate is achieved by properly choosing the parameters $x$, $y$, and $z$ specifying the wave-plate rotations. 
The choice that maximizes the overall success probability of the gate is given by 
\[
|\tan x|= \sqrt{\cot{\left(|\varphi|/2\right)}}, \qquad y=x,
\]
and
\[
\cos z = \sqrt{\cos^4 x +\sin^4 x}.
\]
The four-qubit C$^{3}$Z gate introduces a $\pi$-phase shift if and only if all four qubits are in logical state $|1\rangle$. This gate \cite{Starek2016} is implemented with the help of a two-photon interference 
which occurs at a partially polarizing beam splitter PPBS that is completely transmitting for horizontally polarized photons while it exhibits transmittance $1/3$ for vertically polarized photons 
\cite{Okamoto2005,Langford2005,Kiesel2005}.
Two additional PPBSs  are employed to balance the amplitudes. This probabilistic gate operates in the coincidence basis and its theoretical success probability is $1/9$. 
The auxiliary path qubit is finally removed by keeping only the central output from the second beam displacer. 
This filtration further reduces the success probability of the gate by a factor $\frac{1}{1 + |\sin\varphi|}$.

\subsection*{Quantum state and quantum process tomography}
In our experiment, the output signal and environmental qubits were comprehensively characterized by quantum state tomography. 
We have sequentially projected these two output qubits on the 36 combinations 
of product states $|\psi_j\rangle_S|\psi_k\rangle_E$, where  $|\psi_j\rangle,|\psi_k\rangle \in \mathcal{S}$. This corresponds to measurements in $9$ product two-qubit bases formed 
by all combinations of three mutually unbiased single-qubit bases $\{|0\rangle, |1\rangle\}$, $\{|+\rangle, |-\rangle\}$, $\{|\!\circlearrowright\rangle, |\!\circlearrowleft\rangle\}$.
To avoid problems with absolute calibration of the single-photon detector efficiencies, we do not perform all measurements for a given two-qubit basis simultaneously 
but instead record the number of detection events for each projection sequentially with the same pair of detectors for duration of one second. 
The single-qubit tomograms were obtained by summing the detection events over all measurements on the other qubit, and 
the density matrices of the output signal and environmental qubits were reconstructed from the collected data using the maximum likelihood estimation method \cite{Jezek2003}. 

Similarly, we use quantum process tomography to characterize  the implemented three-qubit CCP gate and the effective decoherence operation $\mathcal{D}$ 
acting on the signal qubit. 
 To reconstruct the operation $\mathcal{D}$, we used the signal qubit quantum state tomography data for all six input signal states from set $\mathcal{S}$,
which represents 36 combinations of input single-qubit state preparations and output single-qubit state projections. 
The complete tomographic characterization of the CCP gate was based on  $6^3\times 6^3=6^6$ combinations of three-qubit product input state preparations and output state projections, 
with inputs given by all products of states from $\mathcal{S}$, $|\psi_j\rangle_P|\psi_k\rangle_S |\psi_l\rangle_E$, and similarly for output projections. 
We optimized the order of input preparations and output projections, or wave plate settings, using a traveling salesman problem solver to minimize 
the required time for wave plates reconfiguration \cite{Hosak2018}.

\begin{figure}[t]
\includegraphics[scale=1]{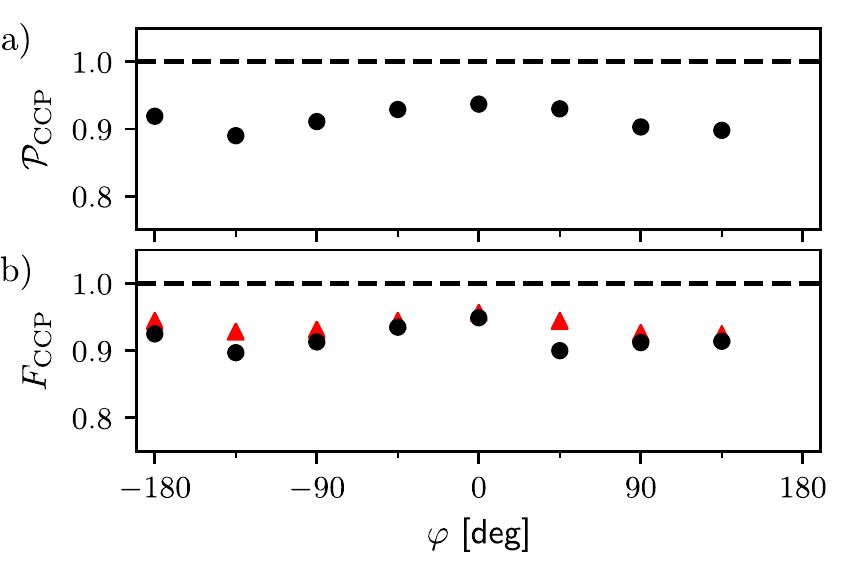}
\centering
\caption{Black dots show the experimentally determined  purity $\mathcal{P}_{\mathrm{CCP}}$ (a) and fidelity $F_{\mathrm{CCP}}$ (b) of the implemented controlled-controlled-phase gate. 
Red triangles show the maximal fidelity that we would obtain after optimal local unitary phase shifts of the output qubits. 
Error bars show three standard deviations and are smaller than the size of the symbols.}
\label{fig:7}
\end{figure}

We represent an $n$-qubit quantum operation $\mathcal{E}$ by its quantum process matrix $\chi$  defined as $\chi= \mathcal{I}\otimes \mathcal{E}(\Phi_n)$, 
where $\mathcal{I}$ denotes the $n$-qubit identity channel and $\Phi_n=|\Phi_n\rangle\langle\Phi_n|$  is a maximally entangled state of $2n$ qubits,
\[
|\Phi_n\rangle=\frac{1}{\sqrt{2^n}}\sum\limits_{i_1, \dots, i_n=0}^1|i_1, \dots, i_n\rangle \otimes |i_1, \dots, i_n\rangle.
\]
The process matrices $\chi$ of $\mathcal{D}$ and of the CCP gate were reconstructed from the experimental data by iterative maximum likelihood estimation algorithm \cite{Jezek2003}. 
We evaluate the quality of the implemented CCP gate by gate fidelity $F_{\mathrm{CCP}}$, defined as 
normalized overlap of the the experimental quantum process matrix $\chi$ with the ideal process matrix of the CCP gate 
$\chi_{\mathrm{CCP}} =|\chi_{\mathrm{CCP}}\rangle\langle\chi_{\mathrm{CCP}}|$, 
\begin{equation}
F_{\mathrm{CCP}}=\frac{\mathrm{Tr}[\chi \chi_{\mathrm{CCP}}]}{\mathrm{Tr}[\chi] \mathrm{Tr}[\chi_{\mathrm{CCP}}]}.
\label{eq:fidelity}
\end{equation}
Here $|\chi_{\mathrm{CCP}}\rangle=\frac{1}{\sqrt{8}}\left[|\Phi_3\rangle+(e^{i\varphi}-1)|111\rangle|111\rangle\right]$.
Besides gate fidelity, we also look at the purity of the implemented operation 
\begin{equation}
\mathcal{P}_{\mathrm{CCP}}=\frac{\mathrm{Tr}[\chi^2]}{\mathrm{Tr}[\chi]^2}.
\label{eq:purity}
\end{equation} 
If $\mathcal{P}_{\mathrm{CCP}}=1$, then the implemented operation is either unitary, or more generally a purity preserving quantum filter 
(a completely positive map with a single Kraus operator). On the other hand, $\mathcal{P}<1$ indicates a noisy operation.

We have performed full quantum process tomography of the CCP gate for eight different coupling strengths $\varphi$ and the resulting gate fidelity and purity are plotted in Fig.~\ref{fig:7}. 
Additionally, we provide a full plot of the reconstructed process matrix for $\varphi = \pi/2$  in Supplementary Material, Fig.~S2. 
The results indicate good and stable performance of the gate for the whole range of coupling strengths $\varphi$, enabling us to demonstrate the studied effect of decoherence suppression.
To estimate the uncertainty of the tomographically obtained quantities, we employed the Monte Carlo method. We used count rates in experimentally obtained tomograms 
as a mean value of a Poisson distribution. For each projection we have drawn 1000 random values to produce 1000 tomograms and reconstructed each of them. 
From the set of the generated results we have then evaluated the standard deviation.

\section*{Acknowledgments}
This work was supported by the Czech Science Foundation ({GA16-17314S}). R. S. acknowledges support by Palack\'{y} University ({IGA-PrF-2018-010} and {IGA-PrF-2019-010}).

\section*{Competing interests}
The authors declare no competing interests.

\section*{Contributions}
R.~F. and J.~F. designed the presented decoherence suppression protocol, J.~F. designed the experimental implementation of the controlled-controlled-phase gate, R.~S. constructed the experimental setup, 
performed the experiment and analyzed data, I.~S. constructed the source of photon pairs, M.~M. and M.~J. supervised the experiment. M.~D. and M.~N. contributed to the development of the utilized experimental techniques. 
R.~F. conceived and supervised the project. All authors contributed to the manuscript writing.


\begin{thebibliography}{0}
\footnotesize


\bibitem{Shor1995}
Shor, P. W. Scheme for reducing decoherence in quantum computer memory. \emph{Phys. Rev. A}~\textbf{52}, R2493 (1995).

\bibitem{Steane1996}
Steane, A. M. Multiple-particle interference and quantum error correction. \emph{Proc. R. Soc. London A}~\textbf{452}, 2551 (1996).

\bibitem{Calderbank1996}
Calderbank, A. R. \& Shor, P. W. Good quantum error-correcting codes exist. \emph{Phys. Rev. A}~\textbf{54}, 1098 (1996).

\bibitem{Nielsen2000}
Nielsen, M.~A. \& Chuang, I.~L. \emph{Quantum Computation and Quantum Information} (Cambridge University Press, Cambridge, 2000.)

\bibitem{Yao2012}
Yao, X.-C. et al. Experimental demonstration of topological error correction. \emph{Nature}~\textbf{482}, 489 (2012).

\bibitem{Bell2014}
Bell,~B.~A. et al. Experimental demonstration of a graph state quantum error-correction code. \emph{Nat. Commun.}~\textbf{5}, 3658 (2014).

\bibitem{Bennett1996}
Bennett, C.~H. el al. Purification of noisy entanglement and faithful teleportation via noisy channels. \emph{Phys. Rev. Lett.}~\textbf{76}, 722 (1996).

\bibitem{Deutsch1996}
Deutsch et al. Quantum privacy amplification and the security of quantum cryptography over noisy channels. \emph{Phys. Rev. Lett.} \textbf{77}, 2818 (1996).

\bibitem{Pan2003}
Pan, J.-W., Gasparoni, S., Ursin, R., Weihs, G. \& Zeilinger, A. Experimental entanglement purification of arbitrary unknown states. \emph{Nature} \textbf{423}, 417 (2003).

\bibitem{Hage2008}
Hage, B. et al. Preparation of distilled and purified continuous-variable entangled states. \emph{Nat. Phys.} \textbf{4}, 915 (2008).

\bibitem{Dong2008}
Dong, R. et al. Experimental entanglement distillation of mesoscopic quantum states. \emph{Nat. Phys.} \textbf{4}, 919 (2008).

\bibitem{Takahashi2010}
Takahashi, H. et al. Entanglement distillation from Gaussian input states. \emph{Nat. Photon.} \textbf{4}, 178 (2010).

\bibitem{Kurochkin2014}
Kurochkin, Y., Prasad, A.~S. \& Lvovsky, A.~I. Distillation of the two-mode squeezed state. \emph{Phys. Rev. Lett.} \textbf{112}, 070402 (2014).

\bibitem{Abdelkhalek2016}
Abdelkhalek D., Syllwasschy, M., Cerf, N.~J., Fiur\'{a}\v{s}ek, J. \& Schnabel, R. Efficient entanglement distillation without quantum memory. \emph{Nat. Commun.} \textbf{7}, 11720 (2016).

\bibitem{Briegel1998}
Briegel, H.~J., D\"{u}r, W., Cirac, J.~I. \& Zoller, P. Quantum repeaters: the role of imperfect local operations in quantum communication. \emph{Phys. Rev. Lett.} \textbf{81}, 5932 (1998).

\bibitem{Duan2001}
Duan, L.-M., Lukin, M.~D., Cirac, J.~I. \& Zoller, P. Long-distance quantum communication with atomic ensembles and linear optics. \emph{Nature} \textbf{414}, 413 (2001).

\bibitem{Chou2007}
Chou, C.-W. et al. Functional quantum nodes for entanglement distribution over scalable quantum networks. \emph{Science} \textbf{316}, 1316 (2007).

\bibitem{Yuan2008}
Yuan, Z.-S. et al. Experimental demonstration of a BDCZ quantum repeater node. \emph{Nature} \textbf{454}, 1098 (2008).

\bibitem{Kwiat2000}
Kwiat, P.~G., Berglund, A.~J., Altepeter, J.~B. \& White, A.~G. Experimental verification of decoherence-free subspaces. \emph{Science} \textbf{290}, 498 (2000).

\bibitem{Kohout2008}
Blume-Kohout, R., Ng, H.~K., Poulin, D. \& Viola, L. Characterizing the structure of preserved information in quantum processes. \emph{Phys. Rev. Lett.} \textbf{100}, 030501 (2008).

\bibitem{Lidar2014}
Lidar, D.~A., Review of decoherence-free subspaces, noiseless subsystems, and dynamical decoupling. \emph{Adv. Chem. Phys.} \textbf{154}, 295 (2014).

\bibitem{Zurek2003}
Zurek, W.-H., Decoherence, einselection, and the quantum origins of the classical. \emph{Rev. Mod. Phys.}~\textbf{75}, 715 (2003).

\bibitem{Hartke2018}
Hartke,~T.~R., Liu,~Y.-Y., Gullans~M.~J. \& Petta,~J.~R. Microwave detection of electron-phonon interactions in a cavity-coupled double quantum dot. \emph{Phys.~Rev.~Lett.}~\textbf{120}, 097701 (2018)

\bibitem{Lisenfeld2015}
Lisenfeld,~J. et al. Observation of directly interacting coherent two-level systems in an amorphous material. \emph{Nat.~Commun.}~\textbf{6}, 6182 (2015)

\bibitem{Rubio-Abadal2018}
Rubio-Abadal,~A. et al. Probing many-body localization in the presence of a quantum bath. \emph{Phys. Rev. X}~\textbf{9}, 041014 (2019). 

\bibitem{Kalb2016}
Kalb,~N. et al. Experimental creation of quantum Zeno subspaces by repeated multi-spin projections in diamond. \emph{Nat.~Commun.}~\textbf{7}, 13111 (2016)

\bibitem{Abobeih2018}
Abobeih,~M.~H. et al. One-second coherence for a single electron spin coupled to a multi-qubit nuclear-spin environment. \emph{Nat.~Commun.}~\textbf{9}, 2552, (2018)

\bibitem{Hernandez-Gomez2018}
Hern\'{a}ndez-G\'{o}mez,~S., Poggiali,~F., Cappellaro,~P. \& Fabbri,~N. Noise spectroscopy of a quantum-classical environment with a diamond qubit. \emph{Phys.~Rev.~B}~\textbf{98}, 214307 (2018)

\bibitem{Gangloff2019}
Gangloff,~D.~A. et al. Quantum interface of an electron and a nuclear ensemble. \emph{Science}~\textbf{364}, 62--66 (2019)

\bibitem{Zhao2011}
Zhao,~N., Wang,~Z.-Y. \& Liu,~R.-B. Anomalous decoherence effect in a quantum bath. \emph{Phys.~Rev.~Lett.}~\textbf{106}, 217205 (2011)

\bibitem{Zwolak2016}
Zwolak,~M., Riedel~C.~J. \& Zurek, W.~H. Amplification, decoherence, and the acquisition of information by spin environments. \emph{Sci.~Rep.}~\textbf{6}, 25277 (2016)

\bibitem{Luitz2017}
Luitz,~D.~J., Huveneers,~F. \& De Roeck,~W. How a small quantum bath can thermalize long localized chains. \emph{Phys.~Rev.~Lett.}~\textbf{119}, 150602 (2017).

\bibitem{Taylor2003}
Taylor,~J.~M., Imamoglu,~A. \& Lukin,~M.~D. Controlling a mesoscopic spin environment by quantum bit manipulation. \emph{Phys.~Rev.~Lett.}~\textbf{91}, 246802 (2003)

\bibitem{Yao2007}
Yao,~W., Liu,~R.-B. \& Sham,~L.~J. Restoring coherence lost to a slow interacting mesoscopic spin bath. \emph{Phys.~Rev.~Lett.}~\textbf{98}, 077602 (2007)

\bibitem{Cappellaro2009}
Cappellaro,~P., Jiang,~L., Hodges,~J.~S. \& Lukin,~M.~D. Coherence and control of quantum registers based on electronic spin in a nuclear spin bath. \emph{Phys.~Rev.~Lett.}~\textbf{102}, 210502 (2009)

\bibitem{Bluhm2010}
Bluhm,~H., Foletti,~S., Mahalu,~D., Umansky,~V. \& Yacoby,~A. Enhancing the coherence of a spin qubit by operating it as a feedback loop that controls its nuclear spin bath. \emph{Phys.~Rev.~Lett.}~\textbf{105}, 216803 (2010)

\bibitem{Kolkowitz2012}
Kolkowitz,~S., Unterreithmeier, Q.~P., Bennett,~S.~D. \& Lukin,~M.~D. Sensing distant nuclear spins with a single electron spin. \emph{Phys.~Rev.~Lett.}~\textbf{109}, 137601 (2012)

\bibitem{Taminiau2012}
Taminiau,~T.~H. et al. Detection and control of individual nuclear spins using a weakly coupled electron spin. \emph{Phys.~Rev.~Lett.}~\textbf{109}, 137602 (2012)

\bibitem{Liu2013}
Liu,~G.-Q., Po,~H.~C., Du,~J., Liu,~R.-B. \& Pan,~X.-Y. Noise-resilient quantum evolution steered by dynamical decoupling. \emph{Nat.~Commun.}~\textbf{4}, 2254 (2013).

\bibitem{Hansom2014}
Hansom,~J. et al. Environment-assisted quantum control of a solid-state spin via coherent dark states. \emph{Nat.~Phys.}~\textbf{10}, 725--730 (2014).

\bibitem{Cramer2016}
Cramer,~J. et al. Repeated quantum error correction on a continuously encoded qubit by real-time feedback. \emph{Nat.~Commun.}~\textbf{7}, 11526 (2016).

\bibitem{Bennett2001}
Bennett, C.~H. et al. Remote state preparation. \emph{Phys. Rev. Lett.}~\textbf{87}, 077902 (2001).

\bibitem{Roszak2015}
Roszak, K., Filip, R. \& Novotn\'{y}, T. Decoherence control by quantum decoherence itself. \emph{Sci. Rep.} \textbf{5}, 9796 (2015).


\bibitem{Korotkov2010}
Korotkov A. N., Keane, K., Decoherence suppression by quantum measurement reversal, \emph{Phys. Rev. A} \textbf{81}, 040103(R) (2010).

\bibitem{Lee2011}
Lee, J.-C. , Jeong, Y.-C.,  Kim, Y.-S., Kim, Y.-H., \emph{Opt. Express} \textbf{19}, 16309 (2011).

\bibitem{Kim2012}
Kim, Y.-S., Lee, J.-C., Kwon, O., Kim,  Y.-H., Protecting entanglement from decoherence using weak measurement and quantum measurement reversal,  \emph{Nature Physics}~\textbf{8}, 117 (2012).


\bibitem{Knill2001}
Knill, E., Laflamme, R. \& Milburn, G.~J. A scheme for efficient quantum computation with linear optics. \emph{Nature}~\textbf{409}, 46 (2001).

\bibitem{Kok2007}
Kok, P. et al. Linear optical quantum computing with photonic qubits. \emph{Rev. Mod. Phys.}~\textbf{79}, 135 (2007).

\bibitem{Jamiolkowski1972}
Jamio\l{}kowsi, A. Linear transformations which preserve trace and positive semidefiniteness of operators. \emph{Rep. Math. Phys.}~\textbf{3}, 275-278 (1972)

\bibitem{Choi1975}
Choi, M.-D. Completely positive linear maps on complex matrices. \emph{Linear Algebra Appl.}~\textbf{10}, 285-290 (1975)

\bibitem{Jezek2003}
Je\v{z}ek, M., Fiur\'{a}\v{s}ek, J. \& Hradil, Z. Quantum inference of states and processes. \emph{Phys. Rev. A}~\textbf{68}, 012305 (2003).

\bibitem{Wooters1998}
Wooters, W.~K. Entanglement of formation of an arbitrary state of two qubits. \emph{Phys. Rev. Lett.}~\textbf{80}, 2245 (1998)

\bibitem{Lv2018}
Lv, D., et al. Simulation of the quantum Rabi model in a trapped ion. \emph{Phys.~Rev.~X}~\textbf{8}, 021027 (2018).

\bibitem{Fluehmann2019}
Fl\"{u}hmann., C., et al. Encoding a qubit in a trapped-ion mechanical oscillator. \emph{Nature}~\textbf{566}, 513-517 (2019).

\bibitem{Yoshikara2019}
Yoshihara, F., et al. Superconducting qubit-oscillator circuit beyond the ultrastrong-coupling regime. \emph{Nat.~Phys.}~\textbf{13}, 44–47 (2017)

\bibitem{Langford2017}
Langford, N. K., et al. Experimentally simulating the dynamics of quantum light and matter at deep-strong coupling. \emph{Nat.~Commun.}~\textbf{8} 1715 (2017).

\bibitem{Braumueller2017}
Braum\"{u}ller, J., et al. Analog quantum simulation of the Rabi model in the ultra-strong coupling regime. \emph{Nat.~Commun.}~\textbf{8} 779 (2017).

\bibitem{LaHaye2009}
LaHaye, M. D., Suh, J., Echternach, P. M., Schwab, K. C. \& Roukes, M. L. Nanomechanical measurements of a superconducting qubit. \emph{Nature}~\textbf{459} 960-964 (2009)

\bibitem{Pirkkalainen2013}
Pirkkalainen, J.-M., Cho, S., Li, J., Paraoanu, G., Hakonen, P. \& Sillanp\"{a}\"{a}, M. Hybrid circuit cavity quantum electrodynamics with a micromechanical resonator. \emph{Nature}~\textbf{494} 211-215 (2013)

\bibitem{Gustafsson2014}
Gustafsson, M. V., et al. Propagating phonons coupled to an artificial atom. \emph{Science}~\textbf{346}, 207-211 (2014).

\bibitem{Starek2018}
St\'{a}rek, R. et al. Experimental realization of SWAP operation on hyper-encoded qubits. \emph{Opt. Express}~\textbf{26}, 8443 (2018).

\bibitem{Fiorentino2004}
Fiorentino, M. \& Wong, F.~N.~C. Deterministic controlled-NOT gate for single-photon two-qubit quantum logic. \emph{Phys. Rev. Lett.}~\textbf{93}, 070502 (2004).

\bibitem{Okamoto2005}
Okamoto, R., Hofmann, H.~F., Takeuchi, S. \& Sasaki, K. Demonstration of an optical quantum controlled-NOT gate without path interference. \emph{Phys. Rev. Lett.}~\textbf{95}, 210506 (2005).

\bibitem{Langford2005}
Langford~N. K. et al. Demonstration of a simple entangling optical gate and its use in Bell-state analysis \emph{Phys. Rev. Lett.}~\textbf{95}, 210504 (2005).

\bibitem{Kiesel2005}
Kiesel, N., Schmid, C., Weber, U., Ursin, R. \& Weinfurter, H. Linear optics controlled-phase gate made simple. \emph{Phys. Rev. Lett.}~\textbf{95}, 210505 (2005).

\bibitem{Lanyon2009}
Lanyon, B.~P. et al. Simplifying quantum logic using higher-dimensional Hilbert spaces. \emph{Nature Phys.}~\textbf{5}, 134 (2009).

\bibitem{Lemr2011}
Lemr, K. et al. Experimental implementation of the optimal linear-optical controlled phase gate. \emph{Phys. Rev. Lett.}~\textbf{106}, 013602 (2011).

\bibitem{Micuda2013}
Mi\v{c}uda, M. et al. Efficient experimental estimation of fidelity of linear optical quantum Toffoli gate. \emph{Phys. Rev. Lett.}~\textbf{111}, 160407 (2013).

\bibitem{Micuda2015}
Mi\v{c}uda, M. et al. Quantum controlled-gate for weakly interacting qubits. \emph{Phys. Rev. A}~\textbf{92}, 022341 (2015).

\bibitem{Starek2016}
St\'{a}rek, R. et al. Experimental investigation of a four-qubit linear-optical quantum logic circuit. \emph{Sci.~Rep.}~\textbf{6}, 33475 (2016).

\bibitem{Hosak2018}
Ho\v{s}\'{a}k, R., St\'{a}rek, R. \& Je\v{z}ek, M. Optimal reordering of measurements for photonic quantum tomography. \emph{Optics Express}~\textbf{26}, 32878 (2018).

\end{thebibliography}
\end{document}